\documentclass{elsart}
\usepackage{graphicx}
\usepackage{amssymb}
\newcommand{\be}{\begin{equation}}
\newcommand{\ee}{\end{equation}}
\newcommand{\bea}{\begin{eqnarray}}
\newcommand{\eea}{\end{eqnarray}}

\newcommand{\eqref}[1]{(\ref{#1})}
\newcommand{\+}{\phantom{-}}

\begin{document}
    
\begin{frontmatter}

\title{Effective string action for 
the U(1)$\times$U(1) dual Ginzburg-Landau theory
beyond the London limit }

\author{Yoshiaki Koma}$^{a,b}$, \ead{ykoma@mppmu.mpg.de}
\author{Miho Koma}$^{c,b}$, \ead{mkoma@mppmu.mpg.de}
\author{Dietmar Ebert}$^{d}$, \ead{debert@physik.hu-berlin.de}
\author{Hiroshi Toki}$^{c}$ \ead{toki@rcnp.osaka-u.ac.jp}

\address{$^{a}$ 
Institute for Theoretical Physics,  Kanazawa University,\\
Kanazawa, Ishikawa 920-1192,  Japan\\
$^{b}$ Max-Planck-Institut f\"ur Physik,
F\"ohringer Ring 6, D-80805 M\"unchen, Germany\\
$^{c}$ Research Center for Nuclear Physics, Osaka University,\\
Mihogaoka 10-1, Ibaraki, Osaka 567-0047, Japan\\
$^{d}$ Institut f\"ur Physik, Humboldt Universit\"at zu Berlin, 
D-10115 Berlin, Germany}

\begin{abstract}
The effective string action of the color-electric flux tube  
in the U(1)$\times$U(1) 
dual Ginzburg-Landau (DGL) theory is studied 
by performing  a path-integral analysis by taking into account
the finite thickness of the flux tube.
The DGL theory, corresponding to the low-energy effective theory of 
Abelian-projected SU(3) gluodynamics, can be expressed as
a [U(1)]$^{3}$ dual Abelian Higgs (DAH) model
with a certain constraint in the Weyl symmetric formulation.
This formulation allows us to adopt quite similar path-integral 
techniques as in the U(1) DAH model, and therefore,
the resulting effective string action in the U(1)$\times$U(1) 
DGL theory has also quite a 
similar structure except the number of color degrees of freedom.
A modified Yukawa interaction appears as a  boundary 
contribution, which is completely due to the finite thickness
of the flux tube, and is reduced into the ordinary Yukawa
interaction in the deep type-II (London) Limit.
\end{abstract}

\end{frontmatter}

\section{Introduction}

The dual superconducting picture of 
the QCD vacuum~\cite{tHooft:1975pu,Mandelstam:1974pi} 
provides an intuitive scenario on the quark confinement 
mechanism.  
The formation of the color-electric flux tube 
can be understood analogously to that of the magnetic Abrikosov 
vortex~\cite{Abrikosov:1957sx,Nielsen:1973ve,Nambu:1974zg}
in an ordinary superconductor  via  the Meissner effect
with the electric Cooper-pair condensation.
In fact, in the QCD vacuum, color-magnetic monopoles are 
condensed, and then, the color-electric field associated 
with quarks is squeezed into an almost one-dimensional object
forming the color-electric flux tube due to the dual
Meissner effect (the dual Higgs mechanism).
The linear rising potential naturally appears 
between the quark and the antiquark, which explains
why a quark cannot be isolated as a free particle.

\par
Remarkably, color-magnetic monopole condensation
is observed by Monte-Carlo simulations of 
lattice gluodynamics in the so-called maximally 
Abelian (MA) 
gauge~\cite{Kronfeld:1987vd,Shiba:1995pu,Shiba:1995db}.
In the MA gauge,  Abelian  dominance~\cite{Ezawa:1982bf,Ezawa:1982ey}
for the nonperturbative quantities, 
such as the string tension~\cite{Suzuki:1990gp,Brandstater:1991sn}
and chiral condensate~\cite{Miyamura:1995xk}, is also observed.
Moreover, it is suggested that  monopole configurations lead
to large contributions to the string tension~\cite{Stack:1994wm}
and chiral condensate~\cite{Miyamura:1995xn}.
These results indicate that Abelian  components of the gluon 
field and monopoles become
relevant degrees of freedom to describe the 
low-energy (long-range) properties of the vacuum.

\par
These numerical results have justified a posteriori
a low-energy effective theory of the gluodynamics 
composed of Abelian degrees of freedom (Abelian-projected
gluodynamics~\cite{tHooft:1981ht}) with monopole condensation.
The resulting effective action is known to have the form
of the Ginzburg-Landau (GL) theory  which in its original 
version would effectively describe bulk properties of the 
ordinary superconductor.
Since now the role of electric and magnetic properties 
is interchanged, this is called the dual Ginzburg-Landau
(DGL) theory~\cite{Maedan:1988yi,Suzuki:1988yq,%
Suganuma:1995ps,Sasaki:1995sa}.
The DGL Lagrangian is composed of the dual gluon field
interacting with the monopole field, and in fact,  
the color-electric flux-tube configuration is obtained 
as the classical solution when the test electric charges
are put into the vacuum.
The thickness of the flux tube is  controlled by the inverse 
masses of the dual gauge boson and the monopole, that is,
the penetration depth $m_{B}^{-1}$
and the coherence length $m_{\chi}^{-1}$.
The latter length characterizes the 
so-called core size of the flux tube.

\par
Although long-distance aspects are dominated by 
Abelian-projected degrees of freedom,  the DGL theory may not be
a complete substitute for all aspects of low-energy gluodynamics,
since there would be an interface in the real theory with
short range physics which is taken care of 
by charged vector and ghost fields.
Such effects would play an important role for the description of 
a dynamical charge density around test charges.
Nevertheless, it is interesting to remind that
the Abelian-projected flux-tube profiles have also been 
studied within the lattice gluodynamics in the MA 
gauge~\cite{Singh:1993jj,Peng:1995yb,Matsubara:1994nq,%
Schilling:1998gz,Gubarev:1999yp,Bornyakov:2001nd},
which have shown quite similar profiles as the 
classical flux-tube solution in the DGL theory.
Therefore it will make sense to regard the DGL 
theory as a starting point of further investigations of the hadronic 
flux tube~\cite{Kamizawa:1993hb,Koma:1999sm,%
Koma:2000wn,Koma:2000hw}
with the pointlike approximation of color-electric test
charges.~\footnote{
The resulting DGL description 
is thus, in particular, justified
for the string away from the charges and 
their dynamical charge clouds.}

\par
From the Abelian-projected SU(2) gluodynamics 
one can construct the U(1) DGL theory 
(often called the U(1) dual Abelian Higgs (DAH) model), where
the  U(1) dual symmetry corresponds to the Cartan subgroup of SU(2). 
One can analyze the physical essence of the dual 
superconducting vacuum and, at the same time,
the structure of the flux tube.
For the description of the dynamics of the flux tube,
it makes sense to 
use a framework  which directly deals with the string-like properties 
of the flux tube. Such framework is 
the {\it string representation} or 
{\it effective string action approach} of the DGL theory, 
which can be obtained by performing a path-integral analysis.
In the U(1) case, based for simplicity,  on the London limit 
approximation $m_{\chi} \to \infty$ ,
the string representation  was
derived in Refs.~\cite{Orland:1994qt,Sato:1995vz,%
Akhmedov:1996mw,Antonov:1998xt}.
The London limit approximation corresponds
to the neglect of the core size of the flux tube.
The resulting leading action 
was shown to be the Nambu-Goto (NG) action with the rigidity 
term~\cite{Polyakov:1986cs,Kleinert:1986bk}.
The string representation of the 
U(1)$\times$U(1) DGL theory
corresponding to the Abelian-projected SU(3) gluodynamics 
was also studied using this 
approximation~\cite{Chernodub:1998ie,Antonov:1998wi}.

\par
Recently, we have considered 
the derivation of the string representation of the U(1) DGL theory 
beyond the London limit, {\it i.e.} we 
have taken into account the 
finite thickness of the flux-tube core
by keeping the mass of the monopole 
finite~\cite{Koma:2001pz}.
The used path-integral analysis combined with differential form 
techniques turned out to provide a very compact and transparent method
for studying these problems including, in particular, also boundary
terms related to the non-confining 
part of the potential.
As a main point, the finite core corrections to the effective string 
action  have been derived.

\par
In this paper, we are going to 
generalize the above work by investigating 
the U(1)$\times$U(1) DGL theory
beyond the London limit.
Note that two of us have shown that the Weyl symmetric formulation
reduces the U(1)$\times$U(1) DGL theory into the sum of
three types of the 
U(1) DAH model, which was quite useful to simplify the 
treatment of the DGL theory and the investigation of 
the flux-tube solution~\cite{Koma:2000wn,Koma:2000hw}.
Therefore, in order to derive the effective string action, 
we pay here special attention to the Weyl symmetric formulation
when using again path-integral techniques in differential form 
notation.

\par
This paper is organized as follows.
In Sec.~\ref{sec:weyl-DGL} 
we introduce the Weyl symmetric formulation 
of the DGL theory in differential form notation. 
In Sec.~\ref{sec:open-string}
the role of the open Dirac string
with  color-electric charge at its ends
for the violation of the dual Bianchi identity 
is considered.
In Sec.~\ref{sec:path-integral} the functional path integration 
in differential form notation is performed
which requires special attention to gauge conditions and 
the Weyl symmetry constraint. This leads to the effective string action,
where the finite thickness of the core of the flux tube is taken into account.
Finally, Sec.~\ref{sec:summary} contains the 
summary.

\section{Weyl symmetric formulation of the DGL theory}
\label{sec:weyl-DGL}

The Weyl symmetric expression of the DGL action in the 
differential form is given by
\bea
S_{\rm DGL} 
= \sum_{i=1}^{3}
\Biggr [
\frac{\beta_{g}}{2}  F_{i}^{2}
+((d-iB_{i})\chi_{i}^{*}, (d+iB_{i})\chi_{i})
+\lambda (|\chi_{i}|^{2}-v^{2})^{2}
\Biggl ],
\label{eqn:weyl-dgl}
\eea
where the ingredients of the DGL theory are
three types of the dual gauge field 
$B_{i}$ ($\sum_{i=1}^{3}B_{i}=0$, $i=1,2,3$), and 
the complex-scalar monopole field 
$\chi_{i}=\phi_{i} \exp (i\eta_{i})$ ($\phi_{i}$, $\eta_{i}$ $\in \Re$,
$i=1,2,3$).
In the presence of external quark sources in the vacuum
describing a static quark-antiquark ($q$-$\bar{q}$) system, 
the dual field strength $F_{i}$ is expressed as 
\be
F_{i} = dB_{i} - 2 \pi * \Sigma_{i}^{(m)},
\ee
with 
\be
*\Sigma_{i}^{(m)} \equiv 
\sum_{j=1}^{3} m_{ij} * \Sigma_{j}^{(e)},
\label{eqn:ele-mag}
\ee
Here $\Sigma_{i}^{(m)}$ ($i=1,2,3$) denotes the 
``color-magnetic'' (Weyl symmetric) representation of 
the color-electric Dirac string $\Sigma_{j}^{(e)}$ ($j=1,2,3$) which 
describes the {\it open} color-electric Dirac string~\cite{Koma:2000hw}.
Let us briefly summarize the definitions of operators 
in differential forms and useful formulae in 
Table~\ref{tbl:note1}, and quote the ingredients of the DGL theory
in Table~\ref{tbl:note2}.
The factor $2\pi m_{ij}$ in front of the 
color-electric Dirac string appears as a result of the extended 
Dirac quantization condition between the color-magnetic charges 
$\vec{Q}_{i}^{(m)} \equiv  g \vec{\epsilon}_{i}$ ($i=1,2,3$) 
and  color-electric charges 
$\vec{Q}_{j}^{(e)}  \equiv e\vec{w}_{j}$ ($j=1,2,3$):
\be
\vec{Q}_{i}^{(m)}  \cdot \vec{Q}_{j}^{(e)} 
= g  \vec{\epsilon}_{i} \cdot e
\vec{w}_{j} =2 \pi m_{ij} \qquad  (eg=4\pi),
\ee
where $\vec{\epsilon}_{i}$ and $\vec{w}_{j}$  
are the root and weight vectors of the 
SU(3) Lie algebra, respectively. 
Explicitly, these vectors are defined as
$\vec{\epsilon}_{1}=(-1/2,\sqrt{3}/2)$, 
$\vec{\epsilon}_{2}=(-1/2,-\sqrt{3}/2)$,
$\vec{\epsilon}_{3}=(1,0)$
and 
$\vec{w}_{1}=(1/2,\sqrt{3}/6)$, $\vec{w}_{2}=(-1/2,\sqrt{3}/6)$,
$\vec{w}_{3}=(0,-1/\sqrt{3})$.
The labels $j=1,2,3$ describe,
for instance, red ($R$), blue ($B$), and  green ($G$) color-electric 
charges, respectively.
The matrix 
\be
m_{ij} \equiv 
2 \vec{\epsilon}_{i} \cdot \vec{w}_{j} = \sum_{k=1}^{3}
\varepsilon_{ijk}
\ee
takes the integer values $\pm 1$ or 0.
Based on these definitions of charges, 
the $R$-$\bar{R}$ system is defined by
$\Sigma_{j=1}^{(e)} = \Sigma \ne 0$,  $\Sigma_{j=2}^{(e)}= 0$,
and $\Sigma_{j=3}^{(e)}=0$, which is also expressed as
$\Sigma_{i=1}^{(m)}=0$, 
$\Sigma_{i=2}^{(m)}=- \Sigma$,
and $\Sigma_{i=3}^{(m)}= \Sigma$
in the color-magnetic representation.
Clearly, the color-electric Dirac string in the color-magnetic 
representation satisfies the relation $\sum_{i=1}^{3} 
\Sigma_{i}^{(m)}=0$ (see, Table~\ref{tbl:string}).

\begin{table}[t]
\caption{Definitions in differential forms in 
four-dimensional 
Euclidean space-time.}
\begin{tabular}{lll}
\hline
$r$-form ($0\le r \le 4$)& $\omega$ &$\omega \equiv   
\frac{1}{r!} \omega_{\mu_1\ldots\mu_r} 
dx_{\mu_1} \wedge \ldots\wedge dx_{\mu_r}$
\\ 
\hline
exterior derivative & $d$ &
 $r$-form $\mapsto$ $(r+1)$-form 
\\
Hodge star & $*$ &
$r$-form $\mapsto$ $(4-r)$-form 
\\
&$**$&multiply by a factor $(-1)^{r}$ for an $r$-form 
\\
codifferential &$\delta \equiv - *d*$&
$r$-form $\mapsto$ $(r-1)$-form 
\\
Laplacian &$\Delta \equiv d\delta + \delta d$&
$r$-form $\mapsto$ $r$-form 
\\
Inner product & $(\omega,\eta)\equiv \int \omega \wedge * \eta$ &
$(\omega,\eta)=
\frac{1}{r!}\int d^{4}x \omega_{\mu_1\ldots\mu_r} 
\eta_{\mu_1\ldots\mu_r} \quad$  ($\omega, \eta \in r$-form)
\\
&&$(\omega)^{2} \equiv (\omega,\omega)$
\\
\hline
\end{tabular}
\label{tbl:note1}
\end{table}

\begin{table}[t]
\caption{Ingredients of the DGL theory in 
differential-form 
notation. The subscripts $i=1,2,3$ denote the label of color 
in the color-magnetic (Weyl symmetric) representation.}
\begin{tabular}{lrl}
\hline
dual gauge field & 
$1$-form &
$B_{i}\equiv B_{i\; \mu} dx_{\mu}$ \\
monopole field  &
$0$-form &
$\chi_{i} \equiv \phi_{i} \exp (i \eta_{i})$ \\
electric Dirac string &
$2$-form &
$\Sigma_{i}^{(m)} \equiv \frac{1}{2}\Sigma_{i\;\mu\nu}^{(m)}
dx_{\mu} \wedge dx_{\nu}$
\\
electric current &
$1$-form &
$j_{i}^{(m)} \equiv j_{i\;\mu}^{(m)} dx_{\mu}$
\\
\hline
\end{tabular}
\label{tbl:note2}
\end{table}

\begin{table}
\caption{The color-electric Dirac string structures in the
color-electric and 
color-magnetic 
representations for the  $q$-$\bar{q}$ 
system, where $\Sigma$ corresponds to the
world sheet of one string singularity.}
\label{tbl:string}
\vspace{0.3cm}

\begin{tabular}{ccccccccc}
\hline
  & & \multicolumn{3}{c}{color-electric rep.} 
  & & \multicolumn{3}{c}{color-magnetic rep.}
\\
\hline
 & & $\+\Sigma_{j=1}^{(e)}$ 
    & $\+\Sigma_{j=2}^{(e)}$  & $\+\Sigma_{j=3}^{(e)}$
 & & $\+\Sigma_{i=1}^{(m)}$ 
    & $\+\Sigma_{i=2}^{(m)}$ & $\+\Sigma_{i=3}^{(m)}$
\\
\cline{3-5}
\cline{7-9}
$R$-$\bar{R}$  
& &$\+\Sigma$  & $\+0$ & $\+0$
& & $\+0$  & $-\Sigma$ & $\+\Sigma$ \\
$B$-$\bar{B}$ 
& &$\+0$ &$\+\Sigma$  &  $\+0$
& &$\+\Sigma$ & $\+0$ & $-\Sigma$\\
$G$-$\bar{G}$  
&  & $\+0$ & $\+0$ &$\+\Sigma$ 
& & $-\Sigma$ & $\+\Sigma$ & $\+0$\\
\hline
\end{tabular}
\end{table}

\par
The physics described by the DGL theory is characterized by 
two mass scales, the masses of the dual gauge boson 
$ m_{B} = \sqrt{2/\beta_{g}}v= \sqrt{3}g v$ and of 
the monopole $m_{\chi}=2\sqrt{\lambda}v$, where 
 $\beta_{g}=2/(3g^{2})$ is the dual gauge coupling 
and $\lambda$  the self-coupling of 
the monopole field.
Recall that their inverses 
correspond to the penetration depth 
and the coherence length, respectively.
We can further define the Ginzburg-Landau (GL) parameter
$\kappa$ as 
\be
\kappa \equiv m_{\chi} / m_{B},
\ee
which controls the vacuum property
of dual superconductivity, 
in particular, whether the vacuum belongs to type-I ($\kappa <1$) 
or type-II ($\kappa >1$) or to their border ($\kappa =1$) .

\section{Open string singularity and the violation of the 
dual Bianchi identity}
\label{sec:open-string}

\par
When we consider the open string system, such as  the 
$R$-$\bar{R}$ system, it is important to pay attention
to the color-electric Dirac string singularity in the 
dual field strength $F_{i}$, since
it is directly related to the violation of the dual Bianchi identity:
\bea
d F_{i} = d^{2}B_{i} -2\pi d* \Sigma_{i}^{(m)} =
+2 \pi * j_{i}^{(m)} \ne 0,
\eea
where $d^{2}B_{i}=0$ and $d*\Sigma_{i}^{(m)}=- * j_{i}^{(m)}$
are used.
Since the latter relation is also written as
\be
\delta \Sigma_{i}^{(m)}  = - j_{i}^{(m)},
\ee
one finds that the color-electric current $j_{i}^{(m)}$ is defined 
as the boundary of the Dirac string $\Sigma_{i}^{(m)}$.
In order to treat the Dirac string singularity further,
it is useful to decompose the dual gauge field $B_{i}$ into two parts
called regular and singular ones,
\be
B_{i}=B_{i}^{\rm reg} + B_{i}^{\rm sing},
\ee
where the singular part  is explicitly written as
\be
B_{i}^{\rm sing} 
= 2 \pi \Delta^{-1}\delta * \Sigma_{i}^{(m)}.
\label{eqn:bsing-def}
\ee
The inverse of the Laplacian, $\Delta^{-1}$, is the Coulomb 
propagator ($\Delta \Delta^{-1}=\Delta^{-1}\Delta =1$). 
Note that $\Delta$ and $\Delta^{-1}$ commute
with the exterior derivative $d$, the codifferential $\delta$, 
and the Hodge star $*$.
Notice further that
$B_{i}^{\rm sing}$ is a background 
gauge field associated with the singular string surface.
By using the relation $d \Delta \delta + \delta \Delta d=1$,
one finds that  Eq.~(\ref{eqn:bsing-def}) leads to 
\be
d B_{i}^{\rm sing} = 2\pi (*\Sigma_{i}^{(m)} + *C_{i}^{(m)}).
\ee
Then, the dual field strength is written as
\be
F_{i}= dB_{i}^{\rm reg}  + 2 \pi  * C_{i}^{(m)},
\ee
where
\be
*C_{i}^{(m)} 
= \Delta^{-1} \delta * j_{i}^{(m)} .
\ee
Clearly, from this expression
we see the violation of the dual Bianchi identity, since
$d^{2}B_{i}^{\rm reg}=0$ and $d*C_{i}^{(m)}=*j_{i}^{(m)}$.
Then, the DGL action is written as
\bea
S_{\rm DGL}
&=& 
\sum_{i=1}^{3}
\Biggr [ \frac{\beta_{g}}{2}
\left ( dB_{i}^{\rm reg} + 2 \pi  * C_{i}^{(m)} \right )^{2}
\nonumber\\*
&&
+(d\phi_{i})^{2}+((B_{i}^{\rm reg}+B_{i}^{\rm sing}
+d \eta_{i}^{\rm reg} ) \phi_{i})^{2}+\lambda (\phi_{i}^{2}-v^{2})^{2}
\Biggl ],
\eea
where the polar decomposition form of the monopole field 
$\chi_{i}=\phi_{i} \exp (i\eta_{i})$  
($\phi_{i}$, $\eta_{i}$ $\in \Re$) is inserted.
Since we are interested in the {\it open}-string system, we 
investigate the case that the phase of the monopole field is
single-valued, so that we have $\eta=\eta^{\rm reg}$.
In general the phase $\eta$ can contain the singular part, which
describes a {\it closed}-string, corresponding
physically to the glueball system \cite{Koma:1999sm}.
Such contribution should be taken into account
when glueballs and quantum 
effects in the dual superconducting vacuum are concerned.

\section{Functional path integration and the effective string action}
\label{sec:path-integral}

\par
The effective string action $S_{\rm eff}(\Sigma_{i}^{(m)})$,
written in terms of 
the color-magnetic representation of the color-electric Dirac string,   
will be obtained after the functional path-integration over both 
the dual gauge fields and the monopole fields.
Explicitly, this can be expressed  by the following relation:
\bea
\mathcal{Z} 
&=&
\int 
\left ( \prod_{i=1}^{3}
\mathcal{D} B_{i}^{\rm reg} \delta [\delta B_{i}^{\rm reg} -f_{i}] \right )
\left ( \prod_{i=1}^{3} \mathcal{D} \phi_{i}^{2} \mathcal{D} 
\eta_{i}^{\rm reg}\right )
\delta [ \sum_{i=1}^{3} \delta B_{i}^{\rm reg} ]
\nonumber\\*
&&
\times 
\exp [-S_{\rm DGL}(B_{i}^{\rm reg},\phi_{i},\eta_{i}^{\rm reg},
\Sigma_{i}^{(m)})]
\nonumber\\*
&=&
\exp [-S_{\rm eff}(\Sigma_{i}^{(m)})].
\eea
It is important to note that,
when we adopt the Weyl symmetric formulation,
 we need to take into account 
the (Lorentz invariant) $\delta$-function constraint
among the dual gauge fields as
\be
\sum_{i=1}^3 \delta B_i^{\rm reg}=0.
\label{eqn:dgf-constraint}
\ee
As a result, the dual gauge symmetry of the DGL theory is kept in 
U(1)$\times$U(1).
In fact,  although the DGL action  (\ref{eqn:weyl-dgl}) 
itself has the [U(1)]$^3$ dual gauge symmetry 
\be
\chi_i \mapsto \chi_i \exp (i \alpha_i),
\quad
B_i^{\rm reg} \mapsto B_i^{\rm reg} -d\alpha_i \quad (i=1,2,3),
\ee
this is finally reduced to U(1)$\times$U(1) 
by  the relation $\sum_{i=1}^3 d\alpha_i=0$, which is 
implied by the constraint (\ref{eqn:dgf-constraint}).
\footnote{Here it may be interesting to comment that in this 
formulation we do not need an additional constraint for the
phase of the monopole field such as $\sum_{i=1}^3 \eta_i = \textit{const}.$
In fact, such a requirement is already implicitly taken into account by 
Eq.~(\ref{eqn:dgf-constraint}).
For instance, the special choice of the gauge, 
(unitary gauge) $\alpha_i \equiv -\eta_i$ ($i=1,2,3$),
leads to the corresponding constraint among the phase 
of the monopole field.}

\par
Let us  perform next the functional path integration over fields.
For this aim, the quadratic term of the dual field strength 
in the action is evaluated as
\bea
\frac{\beta_{g}}{2}
\left ( dB_{i}^{\rm reg} 
+ 2 \pi * C_{i}^{(m)}
\right )^{2}
=
\frac{\beta_{g}}{2}\left ( dB_{i}^{\rm reg} \right )^{2}
+
2 \pi^{2} \beta_{g}\left ( j_{i}^{(m)}, \Delta^{-1}  j_{i}^{(m)} \right ) ,
\eea
where we use
\bea
\left (* C_{i}^{(m)}, * C_{i}^{(m)}  \right )
&=&
\left ( j_{i}^{(m)}, \Delta^{-1}  j_{i}^{(m)} \right ),
\eea
and the cross term vanishes. Next, inserting an identity for 
auxiliary fields $E_{i}$ ($1$-form, $i=1,2,3$),
\bea
&&
\exp \Biggl [  
- ((B_{i}^{\rm reg}+B_{i}^{\rm sing}+d \eta_{i}^{\rm reg} ) \phi_{i})^{2}
\Biggr ] 
\nonumber\\*
&&
=
\phi_{i}^{-4} \int 
\mathcal{D} E_{i} 
\exp \Biggl [  
-
\left \{
\frac{1}{4} (E_{i}, \frac{1}{\phi_{i}^{2}} E_{i})
-i (B_{i}^{\rm reg}+B_{i}^{\rm sing}+d \eta_{i}^{\rm reg}, E_{i})
\right \}
\Biggr ],
\eea
where, in general, $\sum_{i=1}^{3}E_{i} \ne 0$,
and rewriting the $\delta$-function
constraint for the dual gauge fields as a Fourier integral, 
the partition function takes the form
\bea
\mathcal{Z}
&=&
\int 
\mathcal{D} k
\left ( \prod_{i=1}^{3}
\mathcal{D} B^{\rm reg}_{i} \delta [\delta B^{\rm reg}_{i} -f_{B\;i}] \right )
\left ( \prod_{i=1}^{3} \mathcal{D} \phi_{i}^{-2}
\mathcal{D} \eta_{i}^{\rm reg} \right )
\left ( \prod_{i=1}^{3}\mathcal{D} E_{i} \right ) 
\nonumber\\*
&& \!\! \!\! \!\! \!\! \!\! \!\!
\times
\exp \Biggr [
- \sum_{i=1}^{3} \Biggr \{
2 \pi^{2} \beta_{g}\left ( j_{i}^{(m)}, \Delta^{-1}  j_{i}^{(m)} 
\right )+
\frac{\beta_{g}}{2}\left (dB_{i}^{\rm reg}  \right )^{2}
+(d\phi_{i})^{2}
+ \frac{1}{4} (E_{i}, \frac{1}{\phi_{i}^{2}} E_{i})
\nonumber\\*
&& 
-i(B_{i}^{\rm sing},E_{i})
-i (B_{i}^{\rm reg},E_{i}+dk)
-i(d \eta_{i}^{\rm reg}, E_{i})
+\lambda (\phi_{i}^{2}-v^{2})^{2}
\Biggl \}
\Biggl ].
\eea
Here, $k$ is a $1$-form auxiliary field.
Notice that, due to  $(d\eta_{i}^{\rm reg},E_{i})=
(\eta_{i}^{\rm reg},\delta E_{i})$,
the integrations over the phase 
$\eta_{i}^{\rm reg}$ $(i=1,2,3)$ lead to
the constraint $\delta E_{i} =0$ which is resolved by 
introducing the Kalb-Ramond (KR) fields $h_{i}$ (2-form, $i=1,2,3$) as
\be
\delta [\delta E_{i} ]
=
\int \mathcal{D} h_{i} \delta [\delta h_{i} -f_{h\; i}]
\delta [ E_{i} - \delta * h_{i}  ],
\ee
where, in general, $\sum_{i=1}^{3} h_{i} \ne 0$.
Due to the hyper-gauge invariance $h \mapsto h+ d \Lambda$ with a
1-form field $\Lambda$, the hyper-gauge fixing functional also 
appears in order to avoid the overcounting in the integration over $h$.
Then, we have
\bea
\mathcal{Z}
&=&
\int 
\mathcal{D} k
\left ( \prod_{i=1}^{3}\mathcal{D}B^{\rm reg}_{i}
\delta [\delta B_{i}^{\rm reg} -f_{B\;i}] \right )
\left ( \prod_{i=1}^{3}\mathcal{D} \phi_{i}^{-2}\right )
\nonumber\\*
&&
\times
\left ( \prod_{i=1}^{3}\mathcal{D} E_{i} \mathcal{D} h_{i}
\delta [ \delta h_{i} -f_{h\; i}]
\delta [E_{i} - \delta * h_{i} ]
\right ) 
\nonumber\\*
&&
\times
\exp \Biggr [
- \sum_{i=1}^{3} \Biggr \{
2 \pi^{2} \beta_{g}\left ( j_{i}^{(m)}, \Delta^{-1}  j_{i}^{(m)} \right )
+
\frac{\beta_{g}}{2}\left (dB_{i}^{\rm reg} \right )^{2}
+(d\phi_{i})^{2} 
\nonumber\\*
&&
+ \frac{1}{4} (E_{i}, \frac{1}{\phi_{i}^{2}} E_{i})
-i (B_{i}^{\rm sing}, E_{i})
-i (B_{i}^{\rm reg}, E_{i}+dk)
+\lambda (\phi_{i}^{2}-v^{2})^{2}
\Biggl \}
\Biggl ].
\eea
Now the integration over $E_{i}$ $(i=1,2,3)$ leads to
\bea
\mathcal{Z}
&=&
\int 
\mathcal{D} k
\left ( \prod_{i=1}^{3}\mathcal{D}B^{\rm reg}_{i}
\delta [\delta B_{i}^{\rm reg} -f_{B\;i}] \right )
\left ( \prod_{i=1}^{3}\mathcal{D} \phi_{i}^{-2}\right )
\left ( \prod_{i=1}^{3} \mathcal{D} h_{i}
\delta [ \delta h_{i} -f_{h\; i}]
\right ) 
\nonumber\\*
&&
\times
\exp \Biggr [
-  \sum_{i=1}^{3} \Biggr \{
2 \pi^{2} \beta_{g}\left ( j_{i}^{(m)}, \Delta^{-1}  j_{i}^{(m)} \right )
+
\frac{\beta_{g}}{2}\left (dB_{i}^{\rm reg}\right )^{2}
+(d\phi_{i})^{2} 
\nonumber\\*
&&
+ \frac{1}{4} (\delta * h_{i}, 
\frac{1}{\phi_{i}^{2}} \delta * h_{i} )
-i (B_{i}^{\rm sing}, \delta * h_{i}  )
-i (B_{i}^{\rm reg}, \delta * h_{i}+dk)
\nonumber\\*
&&
+\lambda (\phi_{i}^{2}-v^{2})^{2}
\Biggl \}
\Biggl ].
\eea
In order to evaluate the partition function further, we use the relations
\bea
\sum_{i=1}^{3} (B_{i}^{\rm sing},\delta * h_{i} )
&=&
2 \pi \sum_{i=1}^{3} 
(\Delta^{-1}\delta * \Sigma_{i}^{(m)},\delta * h_{i})
\nonumber\\*
&=&
2 \pi \sum_{i=1}^{3} (h_{i}, \Sigma_{i}^{(m)})
+2 \pi \sum_{i=1}^{3} (\delta h_{i},   \Delta^{-1} j_{i}^{(m)} ) ,
\eea
and
\bea
&&
(\delta * h_{i}, \frac{1}{\phi_{i}^{2}} \delta * h_{i})=
(dh_{i},\frac{1}{\phi_{i}^{2}} dh_{i}).
\eea
By inserting an identity 
in form of a path-integration over
$f_{B\;i}$ ,
\be
\textit{const.}= \int \mathcal{D} f_{B\; i} \exp \left [ 
-\frac{\beta_{g}}{2 \xi_{B\;i}} f_{B\;i}^{2} \right ],
\ee
and taking $\xi_{B\;i}=1$, 
we can integrate over $B_{i}^{\rm reg}$ as
\bea
\mathcal{Z}
&=&
\int \mathcal{D} k 
\left ( \prod_{i=1}^{3}\mathcal{D} \phi_{i}^{-2} \mathcal{D} h_{i} 
 \delta [ \delta h_{i} -f_{h\;i}] \right ) 
\nonumber\\*
&&
\times \exp \Biggr [
-  \sum_{i=1}^{3} \Biggr \{
2 \pi^{2} \beta_{g}\left ( j_{i}^{(m)}, \Delta^{-1}  j_{i}^{(m)} 
\right )
\nonumber\\*
&&
+ \frac{1}{2\beta_{g}}
(\delta * h_{i} +dk , \Delta^{-1} (\delta * h_{i} +dk) )
+ (d\phi_{i})^{2}
+ \frac{1}{4} (dh_{i}, \frac{1}{\phi_{i}^{2}} dh_{i}) 
\nonumber\\*
&&
-2 \pi i (h_{i}, \Sigma_{i}^{(m)})
-2 \pi i (\delta h_{i},   \Delta^{-1} j_{i}^{(m)} )  
+\lambda (\phi_{i}^{2}-v^{2})^{2} \Biggl \} \Biggl ],
\eea
where
\bea
(\delta * h_{i}+dk ,\Delta^{-1} (\delta * h_{i}+dk))
=
(h_{i})^{2} - (\delta h_{i}, \Delta^{-1} \delta h_{i}) +(k)^{2}.
\eea
Finally, the integration of $k$ leads to a constant
providing us with the expression
\bea
\mathcal{Z}
&=&
\int
\left ( \prod_{i=1}^{3}\mathcal{D} \phi_{i}^{-2} \mathcal{D} h_{i} 
 \delta [ \delta h_{i} -f_{h\;i}] \right ) 
\nonumber\\*
&&
\times
\exp \Biggr [
-  \sum_{i=1}^{3} \Biggr \{
2 \pi^{2} \beta_{g}\left ( j_{i}^{(m)}, \Delta^{-1}  j_{i}^{(m)} \right )+
\frac{1}{2\beta_{g}}
 (h_{i})^{2} 
 - \frac{1}{2\beta_{g}}(\delta h_{i}, \Delta^{-1} \delta h_{i}) 
\nonumber\\*
&&
+ (d\phi_{i})^{2}
+ \frac{1}{4} (dh_{i}, \frac{1}{\phi_{i}^{2}} dh_{i}) 
-2 \pi i (h_{i}, \Sigma_{i}^{(m)})
-2 \pi i (\delta h_{i},   \Delta^{-1} j_{i}^{(m)} )  
\nonumber\\*
&&
+\lambda (\phi_{i}^{2}-v^{2})^{2}
\Biggl \}
\Biggl ].
\nonumber\\
\eea
As in our previous paper \cite{Koma:2001pz},
we divide the action into three parts as
\be
S = \sum_{i=1}^{3} [S_{i}^{(1)} +S_{i}^{(2)} +S_{i}^{(3)}],
\ee
where 
\bea
S_{i}^{(1)} &=&
2 \pi^2 \beta_g (j_i^{(m)},\Delta^{-1} j_i^{(m)}),
\label{eq:s-coulomb}
\\*
S_{i}^{(2)} &= & 
(d \phi_i)^2 + \frac{1}{4}(dh_i, 
\left \{ \frac{1}{\phi_i^2} - \frac{1}{v^2} \right \} d h_i)
+ \lambda (\phi_i^2 -v^2)^2,\\*
S_{i}^{(3)} &=& 
\frac{1}{4 v^2}(dh_i)^2 + \frac{1}{2 \beta_g} (h_i)^2-
\frac{1}{2 \beta_g} (\delta h_i, \Delta^{-1} \delta h_i)
\nonumber\\*
&&
- 2 \pi i (h_i, \Sigma_i^{(m)}) - 2 \pi i ( \delta h_i,
\Delta^{-1} j_i^{(m)}).
\eea
Since these three actions have completely the same structures 
as in the SU(2) case \cite{Koma:2001pz} except for the labels of $i$, 
the further evaluation of these actions is  straightforward.
The first action $S_{i}^{(1)}$ provides the pure Coulombic 
potential for the static quark-antiquark system, since the 
color-electric currents interact via the Coulomb propagator.
The second and third actions, $S_{i}^{(2)}$ and $S_{i}^{(3)}$,
should be evaluated by taking 
into account the variation of the monopole field $\phi_i$
in the vicinity of the center of the flux tube, which 
defines the ``core'' region of the flux tube.
In fact, its classical profile shows the behavior 
$\phi_i=v$ at the distant region from the center,
while it tends to vanish as $\phi_i \to 0 $ on the Dirac string
if $\Sigma_{i}^{(m)} \ne 0$
(if $\Sigma_i^{(m)} = 0$, the monopole field behaves as
 $\phi_i = v$ everywhere~\cite{Koma:2000wn}).
Effectively, the core region can be 
characterized by the inverse of the monopole 
mass as $\rho \le m_\chi^{-1}$.
It is interesting to note that the second action contributes 
only inside of the core since it vanishes 
in the region where $\phi_i=v$, and its leading 
term contains the NG action with the string tension 
$\sigma^{\rm core}$~\cite{Forster:1974ga}.
At the same time, the third action in the core region 
contributes a boundary contribution $S_{i}^{\rm core} (j_{i}^{(m)})$. 
Then, the action for the core region is 
\bea
S_{i  \; \vert \; <m_{\chi}^{-1}} = 
S_{i}^{\rm core} (j_{i}^{(m)})
+\sigma^{\rm core} \int_{\Sigma_{i}^{(m)}} d^2 \xi_{i} \sqrt{g(\xi_{i})},
\label{eq:s-in}
\eea
where $\xi_{i}^a$ ($a=1,2$) parametrize the string world sheet 
$\tilde{x}_{i\; \mu} (\xi_{i})$,
and $g(\xi_{i})$ is the determinant of the induced metric, 
$g_{ab}(\xi_{i}) \equiv 
\frac{\partial\tilde{x}_{i\;\mu}(\xi_{i})}{\partial \xi_{i}^{a}}
\frac{\partial\tilde{x}_{i\;\mu}(\xi_{i})}{\partial \xi_{i}^{b}}$
~\cite{Forster:1974ga,Davis:1988rw}.
Here, the third action in the outside region $\rho > m_\chi^{-1}$
is responsible for the description of the ``surface'' of the flux tube. 
By inserting a suitable identity in form of a
path-integral over the hyper-gauge fixing function $f_{h\;i}$,
\bea
\textit{const.} &=&
\int \mathcal{D}f_{h\;i} \;
\nonumber\\*
\! &\times &\!\exp \Biggl [ -  \frac{1}{4 v^{2}}\!
\left ([f_{h\;i}+4 \pi v^{2} i D j_{i}^{(m)} ],
D^{-1} \Delta^{-1} [f_{h\;i}+4 \pi v^{2} i Dj_{i}^{(m)}] \right ) \! \Biggr ],
\eea
where $D \equiv (\Delta + m_{B}^{2})^{-1}$ 
is the propagator of the massive KR field,
we can further integrate over the KR field as
\bea
S_{i \; \vert\; >m_{\chi}^{-1}}
&=&
   4 \pi^{2}v^{2}
\left( \Sigma_{i}^{(m)}, D \Sigma_{i}^{(m)} \right) \biggr |_{>m_{\chi}^{-1}}
\nonumber\\*
&&
+2\pi^{2}\beta_{g}  
\left ( j_{i}^{(m)},\left [ D  - \Delta^{-1} \right ] j_{i}^{(m)}
\right ) \biggr |_{>m_{\chi}^{-1}}.
\label{eq:s-out}
\eea
Now ``$|_{>m_{\chi}^{-1}}$'' means that
a corresponding effective cutoff 
should be taken into account.
One finds that the first term describes the interaction between
the color-electric Dirac string via the propagator of 
the massive KR field.
The derivative expansion of the KR propagator 
with respect to the covariant Laplacian on the string world  sheet 
which is defined by
\be
\Delta_{\xi_{i}} = -\frac{1}{\sqrt{g(\xi_{i})}} \partial_{a} g^{ab}(\xi_{i})
\sqrt{g(\xi_{i})}
\partial_{b},
\label{eqn:covariant-Lap}
\ee
leads to the following  form: 
\bea
  4 \pi^{2}v^{2}&&
\left( \Sigma_{i}^{(m)}, D \Sigma_{i}^{(m)} \right)
\biggr |_{>m_{\chi}^{-1}}
\nonumber\\*
=&&
\sigma^{\rm surf}
\int_{\Sigma_{i}^{(m)}} d^{2}\xi_{i} \sqrt{g(\xi_{i})}
 \nonumber\\*
 &&
+
\alpha^{\rm surf}
\int_{\Sigma_{i}^{(m)}} d^{2}\xi_{i} \;   \sqrt{g(\xi_{i})}  g^{ab}(\xi_{i})
\left( \partial_{a} t_{\mu \nu}(\xi_{i}) \right)
  \left( \partial_{b}  t_{\mu \nu}(\xi_{i})\right)
+ \mathcal{O}( \Delta_{\xi_{i}}^{2}),
\label{eqn:expansion}
\eea
where $t_{\mu\nu}(\xi_{i}) = \frac{\epsilon^{ab}}{\sqrt{g(\xi_{i})} }
\frac{\partial \tilde{x}_{\mu}(\xi_{i})}{\partial \xi_{i}^{a}}
\frac{\partial \tilde{x}_{\nu}(\xi_{i})}{\partial \xi_{i}^{b}}$ is an
antisymmetric tensor which determines the orientation of the string
world  sheet $\Sigma_{i}^{(m)}$.
The first term represents the NG action
with the string tension
\be
\sigma^{\rm surf}
= \pi v^{2}\ln \frac{m_B^{2} +m_{\chi}^{2}}{m_B^{2}}
= \pi v^{2}\ln \left(1 +\kappa^{2}\right),
\ee
and  the second term  is the so-called rigidity term with
the negative coefficient
\be
\alpha^{\rm surf}
=
\frac{\pi v^{2} }{2}
\left(\frac{1}{m_{\chi}^{2}+m_B^{2}} - \frac{1}{m_B^{2}}\right)
=
- \frac{\pi \beta_{g}}{4} 
\frac{\kappa^{2}}{1+\kappa^{2}} \qquad (< 0),
\ee
where $\kappa=m_{\chi}/m_B$ is the GL parameter.

\par
Finally, by combining all contributions, the pure Coulomb 
term (\ref{eq:s-coulomb}), the core term (\ref{eq:s-in}), and 
the surface  term (\ref{eq:s-out})
with derivative expansion (\ref{eqn:expansion}),  
we get the effective string action 
\bea
S_{\mathrm{eff}}
&=&
\sum_{i=1}^{3}
\Biggr [
S_{i} (j_{i}^{(m)})
+(\sigma^{\rm core} 
+ \sigma^{\rm surf}) \int_{\Sigma_{i}^{(m)}} d^{2}\xi_{i} \sqrt{g(\xi_{i})}
\nonumber\\*
&&
+ \alpha^{\rm surf}\int_{\Sigma_{i}^{(m)}}
d^{2}\xi_{i} \;   \sqrt{g(\xi_i)}  g^{ab}(\xi_{i})
\left( \partial_{a} t_{\mu \nu}(\xi_{i}) \right)
  \left( \partial_{b} t_{\mu \nu}(\xi_{i})\right)
+ \mathcal{O}( \Delta_{\xi_{i}}^{2})
\Biggl ],
\label{eq:effaction}
\eea
where the boundary (color-electric current) contribution is given by
\bea
S_{i} (j_{i}^{(m)}) &=&S_{i}^{\rm core}(j_{i}^{(m)}) 
+ 2 \pi^{2} \beta_{g} (j_{i}^{(m)}, \Delta^{-1} j_{i}^{(m)}) 
\nonumber\\*
&&
+ 2 \pi^{2} \beta_{g}
\left ( j_{i}^{(m)}, [D  - \Delta^{-1} ] \; j_{i}^{(m)}  \right )
\biggr |_{>m_{\chi}^{-1}}.
\label{effaction-boundary}
\eea
Eqs.~(\ref{eq:effaction}), (\ref{effaction-boundary}) are the main 
results of this paper .

\par
It has to be noted that the boundary term can be simplified only in 
the London limit $m_{\chi}\to \infty$.
In this limit, the first term disappears,
$S_{i}^{\rm core}(j_{i}^{(m)}) \to 0$, and the third term becomes 
an ``exact'' Yukawa minus Coulomb term.
Then, the {\it cancellation of 
the Coulomb term takes place},
and finally the boundary term is reduced into the 
ordinary Yukawa interaction as
\be
S_{i} (j_{i}^{(m)})  \to 
2 \pi^{2} \beta_{g}
\left ( j_{i}^{(m)}, D j_{i}^{(m)}  \right ).
\label{eqn:Yukawa}
\ee
However,  if the vacuum is not in the deep type-II
limit, the boundary contribution has, in principle, a complicated
structure which can only be investigated numerically.

\begin{figure}[t]
\centering
\includegraphics[width=11cm]{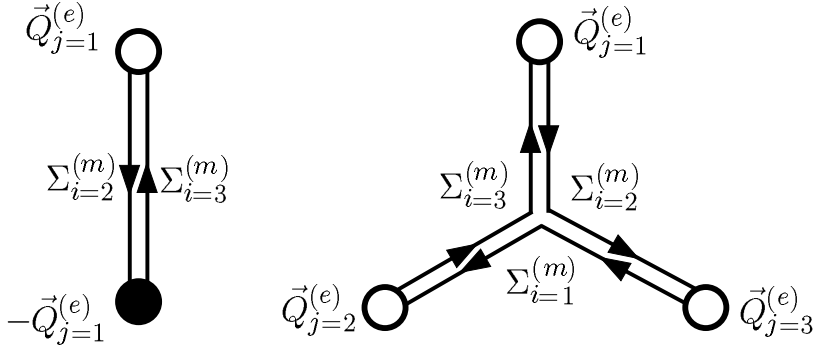}
\caption{The structure of the color-electric Dirac string
in the  color-magnetic representation 
for the mesonic (left) and the baryonic (right) quark  systems.}
\label{fig:string-structure}
\end{figure}

\par
Clearly, the explicit form of the effective string action depends on the 
quark  system, that is to say, the combination of 
the color-electric  charges. In the SU(3) case, one can consider 
not only the  mesonic string, described by
$R$-$\bar{R}$, $B$-$\bar{B}$ and $G$-$\bar{G}$ systems,
but also the {\it baryonic} string, as given by the combination of
three color-electric charges $R$-$B$-$G$ (see, 
Fig.~\ref{fig:string-structure}).
Remarkably, since the form of the action is now the Weyl 
symmetric one, which means invariant under the exchange of the color
labels $i$, one can immediately conclude 
that all mesonic strings do not depend on the color.
The baryonic system is constructed by connecting
the three-colored mesonic strings, $R$-$\bar{R}$, $B$-$\bar{B}$,
and $G$-$\bar{G}$, at a certain {\it junction},  
forming a Y-shaped structure.
This is possible because the sum of all three color charges at
the same point turns out to be zero.
In Ref.~\cite{Chernodub:1998ie}, the static potential among 
three ends is
calculated using the effective string action in the London limit
and its short distance region is described by the Yukawa potential, 
which is consistent with our estimation in Eq.~(\ref{eqn:Yukawa}).
The profile of the baryonic flux-tube solution in the 
DGL theory is investigated
in Refs.~\cite{Kamizawa:1993hb,Koma:2000hw}, where 
also  the  role of the  color-electric Dirac string inside 
the flux tube is clarified.

\section{Summary}\label{sec:summary}

In this paper we have studied the effective string action of 
the color-electric flux tube in the dual Ginzburg-Landau (DGL) 
theory using the path-integral analysis in differential forms, 
thus extending our previous work for the U(1)
dual Abelian Higgs (DAH) model~\cite{Koma:2001pz} to the
U(1)$\times$U(1) case.
By adopting the Weyl symmetric formulation, where the
U(1)$\times$U(1) symmetry is naturally extended to [U(1)]$^{3}$,
the treatment of the DGL theory is found to be rather simple,
since the resulting form is the sum of three types of the 
U(1) DAH model. 
Taking advantage of this feature, we could similarly get the effective 
string action, which is expressed as the sum of the string action 
in the U(1) case. Then the interpretation of the action is 
also straightforward.
The difference is only the number of color-electric charges,
which then allows us to describe not only the mesonic system
but also the baryonic system.
The latter is a typical system in the SU(3) gluodynamics.
In the present paper, we do not describe the
baryon  systems explicitly, which will 
be investigated in a future analysis.

\par
Let us make some comments concerning the form of the 
effective string action in Eqs.~(\ref{eq:effaction}) and 
(\ref{effaction-boundary}).
Analogously to the case of SU(2), the effect of the finite thickness 
of the flux tube leads us again
to a  modified Yukawa interaction as the boundary contribution of 
the open string, which turns 
over to the ordinary Yukawa one only in the London limit.
The derivative expansion of the KR propagator 
with respect to the covariant Laplacian (\ref{eqn:covariant-Lap})
leads to the NG action and the rigidity term as
the leading and next-to-leading parts of the action.
Such an expansion is clearly justified when the fluctuation
of the string world sheet can be treated perturbatively.
If not, one has to handle the expression
of the first term in Eq.~(\ref{eq:s-out}) directly.

\par
Finally, it is interesting to note that we can refer to the
interaction among string world sheets of
various colors without solving their dynamics.
Since the sum of two types of color charges
reduces to another type of {\it anti}-color charge, 
for instance as $R+B \to \bar{G}$, one finds that
if two strings composed of $R$-$\bar{R}$ and $B$-$\bar{B}$ 
interact with each other, this will be 
equivalent to a $\bar{G}$-$G$ string.
Then, it is easy to imagine that the force between them
must be attractive.

\par
One of the authors (Y.K.) is partially supported  by 
the Ministry of Education, Science, Sports and Culture,
Japan, Grant-in-Aid for Encouragement of Young 
Scientists (B), 14740161, 2002.
We would like to thank 
D. Antonov  and E.-M. Ilgenfritz for useful discussions.


\begin{thebibliography}{10}

\bibitem{tHooft:1975pu}
G.~'t~Hooft,
\newblock in {\em High-Energy Physics. Proceedings of the EPS International
  Conference, Palermo, Italy, 23-28 June 1975, Vol. 2}, edited by A.~Zichichi,
  pp. 1225--1249, Bologna, 1976.

\bibitem{Mandelstam:1974pi}
S.~Mandelstam,
\newblock Phys. Rept. {\bf 23C}, 245 (1976).

\bibitem{Abrikosov:1957sx}
A.~A. Abrikosov,
\newblock Sov. Phys. JETP {\bf 5}, 1174 (1957).

\bibitem{Nielsen:1973ve}
H.~B. Nielsen and P.~Olesen,
\newblock Nucl. Phys. {\bf B61}, 45 (1973).

\bibitem{Nambu:1974zg}
Y.~Nambu,
\newblock Phys. Rev. {\bf D10}, 4262 (1974).

\bibitem{Kronfeld:1987vd}
A.~S. Kronfeld, G.~Schierholz, and U.~J. Wiese,
\newblock Nucl. Phys. {\bf B293}, 461 (1987).

\bibitem{Shiba:1995pu}
H.~Shiba and T.~Suzuki,
\newblock Phys. Lett. {\bf B343}, 315 (1995), hep-lat/9406010.

\bibitem{Shiba:1995db}
H.~Shiba and T.~Suzuki,
\newblock Phys. Lett. {\bf B351}, 519 (1995), hep-lat/9408004.

\bibitem{Ezawa:1982bf}
Z.~F. Ezawa and A.~Iwazaki,
\newblock Phys. Rev. {\bf D25}, 2681 (1982).

\bibitem{Ezawa:1982ey}
Z.~F. Ezawa and A.~Iwazaki,
\newblock Phys. Rev. {\bf D26}, 631 (1982).

\bibitem{Suzuki:1990gp}
T.~Suzuki and I.~Yotsuyanagi,
\newblock Phys. Rev. {\bf D42}, 4257 (1990).

\bibitem{Brandstater:1991sn}
F.~Brandstaeter, U.~J. Wiese, and G.~Schierholz,
\newblock Phys. Lett. {\bf B272}, 319 (1991).

\bibitem{Miyamura:1995xk}
O.~Miyamura,
\newblock Nucl. Phys. Proc. Suppl. {\bf 42}, 538 (1995).

\bibitem{Stack:1994wm}
J.~D. Stack, S.~D. Neiman, and R.~J. Wensley,
\newblock Phys. Rev. {\bf D50}, 3399 (1994), hep-lat/9404014.

\bibitem{Miyamura:1995xn}
O.~Miyamura,
\newblock Phys. Lett. {\bf B353}, 91 (1995).

\bibitem{tHooft:1981ht}
G.~'t~Hooft,
\newblock Nucl. Phys. {\bf B190}, 455 (1981).

\bibitem{Maedan:1988yi}
S.~Maedan and T.~Suzuki,
\newblock Prog. Theor. Phys. {\bf 81}, 229 (1989).

\bibitem{Suzuki:1988yq}
T.~Suzuki,
\newblock Prog. Theor. Phys. {\bf 80}, 929 (1988).

\bibitem{Suganuma:1995ps}
H.~Suganuma, S.~Sasaki, and H.~Toki,
\newblock Nucl. Phys. {\bf B435}, 207 (1995), hep-ph/9312350.

\bibitem{Sasaki:1995sa}
S.~Sasaki, H.~Suganuma, and H.~Toki,
\newblock Prog. Theor. Phys. {\bf 94}, 373 (1995).

\bibitem{Singh:1993jj}
V.~Singh, D.~A. Browne, and R.~W. Haymaker,
\newblock Phys. Lett. {\bf B306}, 115 (1993), hep-lat/9301004.

\bibitem{Peng:1995yb}
Y.-c. Peng and R.~W. Haymaker,
\newblock Phys. Rev. {\bf D52}, 3030 (1995), hep-lat/9503015.

\bibitem{Matsubara:1994nq}
Y.~Matsubara, S.~Ejiri, and T.~Suzuki,
\newblock Nucl. Phys. Proc. Suppl. {\bf 34}, 176 (1994), hep-lat/9311061.

\bibitem{Schilling:1998gz}
K.~Schilling, G.~S. Bali, and C.~Schlichter,
\newblock Nucl. Phys. Proc. Suppl. {\bf 73}, 638 (1999), hep-lat/9809039.

\bibitem{Gubarev:1999yp}
F.~V. Gubarev, E.-M. Ilgenfritz, M.~I. Polikarpov, and T.~Suzuki,
\newblock Phys. Lett. {\bf B468}, 134 (1999), hep-lat/9909099.

\bibitem{Bornyakov:2001nd}
V.~Bornyakov {\em et~al.},
\newblock Nucl. Phys. Proc. Suppl. {\bf 106}, 634 (2002), hep-lat/0111042.

\bibitem{Kamizawa:1993hb}
S.~Kamizawa, Y.~Matsubara, H.~Shiba, and T.~Suzuki,
\newblock Nucl. Phys. {\bf B389}, 563 (1993).

\bibitem{Koma:1999sm}
Y.~Koma, H.~Suganuma, and H.~Toki,
\newblock Phys. Rev. {\bf D60}, 074024 (1999), hep-ph/9902441.

\bibitem{Koma:2000wn}
Y.~Koma and H.~Toki,
\newblock Phys. Rev. {\bf D62}, 054027 (2000), hep-ph/0004177.

\bibitem{Koma:2000hw}
Y.~Koma, E.-M. Ilgenfritz, T.~Suzuki, and H.~Toki,
\newblock Phys. Rev. {\bf D64}, 014015 (2001), hep-ph/0011165.

\bibitem{Orland:1994qt}
P.~Orland,
\newblock Nucl. Phys. {\bf B428}, 221 (1994), hep-th/9404140.

\bibitem{Sato:1995vz}
M.~Sato and S.~Yahikozawa,
\newblock Nucl. Phys. {\bf B436}, 100 (1995), hep-th/9406208.

\bibitem{Akhmedov:1996mw}
E.~T. Akhmedov, M.~N. Chernodub, M.~I. Polikarpov, and M.~A. Zubkov,
\newblock Phys. Rev. {\bf D53}, 2087 (1996), hep-th/9505070.

\bibitem{Antonov:1998xt}
D.~Antonov and D.~Ebert,
\newblock Eur. Phys. J. {\bf C8}, 343 (1999), hep-th/9806153.

\bibitem{Polyakov:1986cs}
A.~M. Polyakov,
\newblock Nucl. Phys. {\bf B268}, 406 (1986).

\bibitem{Kleinert:1986bk}
H.~Kleinert,
\newblock Phys. Lett. {\bf B174}, 335 (1986).

\bibitem{Chernodub:1998ie}
M.~N. Chernodub and D.~A. Komarov,
\newblock JETP Lett. {\bf 68}, 117 (1998), hep-th/9809183.

\bibitem{Antonov:1998wi}
D.~Antonov and D.~Ebert,
\newblock Phys. Lett. {\bf B444}, 208 (1998), hep-th/9809018.

\bibitem{Koma:2001pz}
Y.~Koma, M.~Koma, D.~Ebert, and H.~Toki,
\newblock JHEP {\bf 08}, 047 (2002), hep-th/0108138.


\bibitem{Forster:1974ga}
D.~Foerster,
\newblock Nucl. Phys. {\bf B81}, 84 (1974).

\bibitem{Davis:1988rw}
R.~L. Davis and E.~P.~S. Shellard,
\newblock Phys. Lett. {\bf B214}, 219 (1988).

\end{thebibliography}

\end{document}